%Paper: 9204219
%From: manohar@sphal.ucsd.edu
%Date: Mon, 13 Apr 92 11:43:38 PDT

%
% INSTRUCTIONS FOR TeXING PAPER
%
% uses macro file harvmac.tex
%
\input harvmac
%%%%%%%%%%%%%%%%%%%%%%%%%%%%%%%%%%%%%%%%%%%%%%%%%%%%%%%%%%%%%%%%%%%%%%
%
%  UCSD macros to overwrite some of the definitions in harvmac.tex
%  (include after harvmac.tex)
%  last modified 4/92
%
%%%%%%%%%%%%%%%%%%%%%%%%%%%%%%%%%%%%%%%%%%%%%%%%%%%%%%%%%%%%%%%%%%%%%%%
%
% modify the output routine for the little format
%
\ifx\answ\bigans
\else
\output={
  \almostshipout{\leftline{\vbox{\pagebody\makefootline}}}\advancepageno
}
\fi
%
%
% address
%
\def\mayer{\vbox{\sl\centerline{Department of Physics 0319}%
\centerline{University of California, San Diego}
\centerline{9500 Gilman Drive}
\centerline{La Jolla, CA 92093-0319}}}
%
% grant numbers
%

%
% preprint number
%
\def\UCSD#1#2{\noindent#1\hfill #2%
\bigskip\supereject\global\hsize=\hsbody%
\footline={\hss\tenrm\folio\hss}}% restores pagenumbers
%
% abstract
%
\def\abstract#1{\centerline{\bf Abstract}\nobreak\medskip\nobreak\par #1}
%
%
% titlefont
%
%
\edef\tfontsize{ scaled\magstep3}
 \tfontsize  \tfontsize
 \tfontsize \font\titlei=cmmi10 \tfontsize
\font\titleis=cmmi7 \tfontsize \font\titleiss=cmmi5 \tfontsize
\font\titlesy=cmsy10 \tfontsize \font\titlesys=cmsy7 \tfontsize
\font\titlesyss=cmsy5 \tfontsize  \tfontsize
\skewchar\titlei='177 \skewchar\titleis='177 \skewchar\titleiss='177
\skewchar\titlesy='60 \skewchar\titlesys='60 \skewchar\titlesyss='60
%
%\def\titlefont{\def\rm{\fam0\titlerm}% switch to title font
%\textfont0=\titlerm \scriptfont0=\titlerms \scriptscriptfont0=\titlermss
%\textfont1=\titlei \scriptfont1=\titleis \scriptscriptfont1=\titleiss
%\textfont2=\titlesy \scriptfont2=\titlesys \scriptscriptfont2=\titlesyss
%\textfont\itfam=\titleit \def\it{\fam\itfam\titleit}\rm}
%
%
% math symbols
%
%---------------------------------------------------------------------
%
\def\inv{^{\raise.15ex\hbox{${\scriptscriptstyle -}$}\kern-.05em 1}}
  %prime
\def\lbar{{\lower.35ex\hbox{$\mathchar'26$}\mkern-10mu\lambda}} %lambda bar

%
%
% various slashed symbols
%
%
 % slashes a character
\def\dsl{\,\raise.15ex\hbox{/}\mkern-13.5mu D} %this one can be subscripted
\def\delsl{\raise.15ex\hbox{/}\kern-.57em\partial}
\def\Ksl{\hbox{/\kern-.6000em\rm K}}
\def\Asl{\hbox{/\kern-.6500em \rm A}}
\def\Dsl{\hbox{/\kern-.6000em\rm D}} %roman D
\def\Qsl{\hbox{/\kern-.6000em\rm Q}}
\def\gradsl{\hbox{/\kern-.6500em$\nabla$}}
%
% space and backspace in l mode
%
\def\lspace{\ifx\answ\bigans{}\else\qquad\fi}
\def\lbspace{\ifx\answ\bigans{}\else\hskip-.2in\fi} % $$\lbspace...$$
%
%     boxes an equation
%
\def\boxeqn#1{\vcenter{\vbox{\hrule\hbox{\vrule\kern3pt\vbox{\kern3pt
        \hbox{${\displaystyle #1}$}\kern3pt}\kern3pt\vrule}\hrule}}}
%
%     draw a little box (end of proof symbol)
%     e.g. \mbox{.1}{.1}
%
\def\mbox#1#2{\vcenter{\hrule \hbox{\vrule height#2in
\kern#1in \vrule} \hrule}}
%
%
%
%     curly letters
%
   %curly letters

   \def\CL{{\cal L}}
\def\CM{{\cal M}}  \def\CO{{\cal O}}

%
%
%
%     derivatives
%
%

%

\def\bar#1{\overline{#1}}

\def\bra#1{\left\langle #1\right|}
\def\ket#1{\left| #1\right\rangle}

\def\vector#1{{\vec{#1}}}

\def\darr#1{\raise1.5ex\hbox{$\leftrightarrow$}\mkern-16.5mu #1}

%
 %pound sterling
%
 %puts a small half in a displayed eqn
\def\frac#1#2{{\textstyle{#1\over #2}}} %puts a small fraction
%in a displayed eqn
%
%
%     various math operators
%
%

%
%
%
%

%
%       relations
%
\def\ltap{\ \raise.3ex\hbox{$<$\kern-.75em\lower1ex\hbox{$\sim$}}\ }
\def\gtap{\ \raise.3ex\hbox{$>$\kern-.75em\lower1ex\hbox{$\sim$}}\ }
\def\gl{\ \raise.5ex\hbox{$>$}\kern-.8em\lower.5ex\hbox{$<$}\ }
\def\roughly#1{\raise.3ex\hbox{$#1$\kern-.75em\lower1ex\hbox{$\sim$}}}
%
%
%       This defines et al., i.e., e.g., cf., etc.
\def\ie{\hbox{\it i.e.}}        
        
\def\etal{\hbox{\it et al.}}

\def\np#1#2#3{{Nucl. Phys. } B{#1} (#2) #3}
\def\pl#1#2#3{{Phys. Lett. } {#1}B (#2) #3}
\def\prl#1#2#3{{Phys. Rev. Lett. } {#1} (#2) #3}
\def\physrev#1#2#3{{Phys. Rev. } {#1} (#2) #3}
\def\ap#1#2#3{{Ann. Phys. } {#1} (#2) #3}

\relax

\def\lqcd{\Lambda_{\rm QCD}}
\def\qqbar{Q\bar Q}
\def\rq{\Lambda_Q}
\def\rqinv{r_Q^{-1}}
\def\rqb{r_B}
\def\rqbinv{\rqb^{-1}}
\def\mev{{\rm MeV}}
\def\gev{{\rm GeV}}
\def\pseudo{P_v}
\def\vector{V_v^\mu}
\def\ce{c_E}
\def\cb{c_B}

\centerline{{\titlefont{A QCD Calculation of}}}
\bigskip
\centerline{{\titlefont{the Interaction of
Quarkonium with Nuclei}}}
\bigskip
\vskip .5in
\centerline{Michael Luke, Aneesh V. Manohar, and Martin J.
Savage}
\bigskip\medskip
\mayer
\bigskip\bigskip
\vfill
\abstract{
The interaction of quarkonium with nuclei is studied in the
$m_Q\rightarrow \infty$ limit of QCD, where the binding energy is found
to be exactly computable.  The dominant contribution to the interaction
is from two-gluon operators. The forward matrix elements of these
two-gluon operators can be determined from the QCD scale anomaly, and
from deep inelastic scattering.  We apply our results to the $\Upsilon$
and $J/\psi$, treating the $\qqbar$ interaction as purely Coulombic.  We
find the $\Upsilon$ binds in nuclear matter with a binding energy of a
few $\mev$, while for the $J/\psi$ binding is of order 10 $\mev$.  For
the $J/\psi$ in particular we expect confinement effects to produce
large corrections to this result.  }
\vfill
%\draftmode
\UCSD{UCSD/PTH 92-12}{March 1992}

There has been some recent interest in the possibility of bound states
occurring between quarkonium and nuclei
\ref\brod1{S.J. Brodsky, I. Schmidt and G.F. de Teramond,
\prl{64}{1990}{1011}.}
\ref\dave{D.A. Wasson, \prl{67}{1991}{2237}.}.
By scaling $K$-nucleon scattering from high energies and modifying the
interaction due to the reduced size of charmonium the binding energy of
the $\eta_c$ in nuclear matter \dave\ is estimated to be $\sim 30$ MeV.
The motivation for studying the interactions between quarkonium and
nuclei is clear. The interactions of nuclei cannot be computed starting
directly from the QCD Lagrangian. However, they can be parameterised by
using a phenomenological Lagrangian with nucleon and meson degrees of
freedom. Thus a study of nuclear forces provides some understanding of
the dynamics of nucleons and the pions which are produced by the
spontaneous breaking of chiral symmetry. The interactions of quarkonium
allows us to study a different aspect of QCD, the interactions of
gluons. The most significant difference between the nucleon-nucleon
potential and a quarkonium-nucleon potential is that there is no valence
light quark exchange in the latter. The coupling constants of the
interactions between quarkonium with pions, nucleons and other light
states arise solely from gluon induced interactions.

In this paper, we study the interactions of a $\qqbar$ bound state with
nuclear matter, in the limit that the mass $m_Q$ of the quark, and
therefore the inverse radius $\rqinv\sim
\alpha_s\left(\rqinv\right)m_Q$ of the
$\qqbar$ bound state, is much larger than the QCD scale $\lqcd$.  We
then apply this to the cases of experimental interest, the $\bar c c$
and $\bar b b$ bound states, and calculate their binding energies in
nuclear matter.

At scales $\mu\ll\rqinv$, the $\qqbar$ bound state looks like a colour
singlet, pointlike object.  Its dynamics will be described at low
momenta by an effective Lagrangian in which non-renormalizable terms are
suppressed by powers of the compositeness scale $\rq=\rqinv$
($\simeq\alpha(\rq)m_Q$ for a Coulomb bound state).  The gauge invariant
gluon couplings of lowest dimension are the dimension 7
operators\foot{This is a just a covariant formulation of the leading
terms in the usual multipole expansion\ref\multipole{K.~Gottfried,
\prl{40}{1978}{538}\semi T.M.~Yan, \physrev{D22}{1980}{1652}
\semi M.~Voloshin, \np{154}{1979}{365}.}.}
\eqn\lag{\eqalign{
\CL_{\rm int} = \sum_v{1\over\rq^3}\left(P^\dagger_v P_v - V_{\mu v}^\dagger
V^\mu_v\right)&\left(\ce \CO_E+\cb \CO_B\right),\cr
\CO_E \equiv - G^{\mu\alpha A} G_{\alpha }^{\nu A}v_\mu v_\nu,\qquad
\CO_B  &\equiv {1\over 2}  G^{\alpha \beta A} G_{\alpha\beta}^A
-G^{\mu\alpha A} G_{\alpha }^{\nu A}v_\mu v_\nu, }} where $\pseudo$ and
$\vector$ create pseudoscalar ($\eta_c$ or $\eta_b$) and vector
($J/\psi$ or $\Upsilon$) mesons respectively, with four velocity
$v^\mu$.  All the sensitivity to the detailed structure of the meson is
encoded in the coupling constants $\ce$ and $\cb$, which must be
calculated from the underlying QCD dynamics.

Some explanation of \lag\
is in order.  Since the interaction of a heavy state with light degrees
of freedom does not change the velocity of the heavy state in the
$m_Q\rightarrow\infty$ limit, the velocity is conserved, and it is
convenient to describe mesons with different velocities by distinct
fields\ref\georgi{H.~Georgi, \pl{240}{1990}{447}.}.  The fields
$\vector$ and $\pseudo$ are related to the standard fields $P$ and
$V^\mu$ by the field redefinitions
\eqn\redefn{\vector(x)=\sqrt{2m}e^{im v_\alpha x^\alpha}V^\mu(x),\qquad
\pseudo(x)=\sqrt{2m}e^{im v_\alpha x^\alpha}P(x),}
where $m$ is the meson mass, and physical vector states satisfy
$v_\mu\vector = 0$.  Writing the momentum as
\eqn\momentum{p^\mu=mv^\mu+k^\mu,}
we see that derivatives acting on the redefined fields bring down
factors of the small ``residual'' momentum $k^\mu$, so the derivative
expansion does not contain potentially troublesome terms proportional to
$mv^\mu/\rq$.  Also, in the heavy quark limit, the quark spin decouples
and the resulting spin symmetry \ref\wise{N. Isgur and M.B. Wise
\pl{232}{1989}{113};
\pl{237}{1990}{527}.} forces the masses and couplings of the
pseudoscalar and vector bound states to be equal (the minus sign in
\lag\ is because a physical vector particle has spacelike
polarization). There is also no term which converts a pseudoscalar to a
vector at leading order in $1/m_Q$.  Hence, in the
$m_Q\rightarrow\infty$ limit, the binding to nuclei of a vector meson is
equal to that of the corresponding pseudoscalar meson.

In the rest frame of the $\qqbar$ bound state, $v=(1,0,0,0)$, and the
operators $\CO_E$ and $\CO_B$ become
\eqn\restlag{
\CO_E =  {\bf E^A}\cdot
{\bf E^A} ,\qquad \CO_B= {\bf B^A}\cdot{\bf B^A}, } where ${\bf E^A}$
and ${\bf B^A}$ are the colour electric and magnetic fields.  (This
explains the choice of operators and coefficients in Eq.~\lag.) The
coefficients $\ce$ and $\cb$ can be obtained by computing the energy
shift of the $\qqbar$ bound state in an external electric and magnetic
field, \ie\ by computing the quadratic Stark and Zeeman energies. The
dominant contribution is the quadratic Stark effect, since magnetic
effects are suppressed by a factor of $\vec{v} /c\approx \alpha_s(\rq)$
relative to electric effects for each replacement $E\rightarrow B$.

The operators $\CO_E$ and $\CO_B$ can be written as linear combinations
of the gluon energy-momentum tensor,
\eqn\emtensor{
T^{\mu\nu}_{\rm gluon} = {1\over 4} g^{\mu \nu} G^{\alpha\beta A}
G^A_{\alpha\beta} - G^{\mu\alpha A} G^{\nu A}_\alpha, } and the
divergence of the scale current,\foot{For simplicity, we will work in
the limit that the light quark masses are zero. It is straightforward to
include light quark mass effects \ref\annalspaper{R.S.~Chivukula, \etal,
\ap{192}{1989}{93}.}.  The trick of using the scale anomaly to compute
the matrix element of $G^2$ is due to Voloshin and
Zakharov\ref\voloshin{M.~Voloshin and V.~Zakharov,
\prl{45}{1980}{688}.}.}
\eqn\scaleanom{
\partial^\mu s_\mu = T^\alpha_\alpha = {\beta(g)\over 2g}
G^{\alpha\beta A}G^A_{\alpha\beta}.  } Note that the energy-momentum
tensor $T^{\mu\nu}$ that occurs in Eq.~\scaleanom\ is the full
energy-momentum tensor, whereas $T^{\mu\nu}_{\rm gluon}$ in
Eq.~\emtensor\ is only the gluon contribution to the energy-momentum
tensor. The expressions for $\CO_E$ and $\CO_B$ are
\eqn\coexp{\eqalign{
\CO_E &= T^{\mu\nu}_{\rm gluon} v_\mu v_\nu - {g\over 2 \beta(g)}
\ T^\alpha_\alpha ,\cr
\CO_B &= T^{\mu\nu}_{\rm gluon} v_\mu v_\nu + {g\over 2\beta(g)}
\ T^\alpha_\alpha .\cr
}} The operators $\CO_E$ and $\CO_B$ in the effective Lagrangian
Eq.~\lag\ are renormalised at the scale $\rq$, and the operator
$T^{\mu\nu}_{\rm gluon}$ in Eq.~\coexp\ is also renormalised at $\rq$.
(The operator $T^\alpha_\alpha$ is scale independent.)  The QCD beta
function at the scale $\rq$ can be written as
\eqn\betafn{
\mu {d\over d\mu} g = \beta(g) = - b_Q {g^3\over 16 \pi^2} + \ldots,
} so that Eq.~\coexp\ can be rewritten as
\eqn\coexpnew{\eqalign{
\CO_E &= T^{\mu\nu}_{\rm gluon} v_\mu v_\nu + {2\pi \over b_Q \alpha_s(\rq)}
\ T^\alpha_\alpha ,\cr
\CO_B &= T^{\mu\nu}_{\rm gluon} v_\mu v_\nu -
{2\pi \over b_Q \alpha_s(\rq)}\ T^\alpha_\alpha .\cr }} The gluon
operator in Eq.~\lag\ can be rewritten in the form
\eqn\oeobnew{
c_E O_E + c_B O_B =\left(c_E+c_B\right)\left( T^{\mu\nu}_{\rm gluon}
v_\mu v_\nu\right) +
\left(c_E-c_B\right){2\pi \over b_Q \alpha_s(\rq)}
T^\alpha_\alpha.  } The matrix elements of $T^{\mu\nu}_{\rm gluon}$ and
$T^\alpha_\alpha$ in a nucleon state are known at zero momentum
transfer,
\eqn\zeromom{\eqalign{
\bra{p} T^{\alpha}_\alpha \ket{p} &= 2 M^2,\cr
\bra{p} T^{\mu\nu}_{\rm gluon} \ket{p} &=
2V_2(\mu)
\left(p^\mu p^\nu -
\frac 14 g^{\mu\nu} p^2\right),
}} where $M$ is the nucleon mass. The value of $V_2$, the gluon momentum
fraction in the nucleon, is measured in deep inelastic scattering, and
its value can be extracted from the $F_1$ (or $F_2$) structure function.
The usual value quoted in the literature is $V_2(\mu^2 = 16\ \gev^2)
=0.44$ \ref\field{See for example R.D.~Field,
Applications of Perturbative QCD, Addison-Wesley Publishing
Company, 1989.}.  The $\mu$ dependence of $V_2$ may be found in \field.

The forward scattering amplitude of a $\qqbar$ meson
off nucleons is completely determined in terms of the measured value
of $V_2$,
\eqn\fscat{\CM_{\rm fwd}
=2 V_2(\rq)\left({\ce+\cb\over\rq^3}\right)M^2\left( \gamma^2 -
\frac 14 \right) +
2 M^2 \left({\ce-\cb\over\rq^3}\right){2\pi \over b_Q \alpha_s(\rq)},}
where $\gamma$  ($=v^0$)
is the time dilation factor in the rest frame
of the nucleon. The forward scattering amplitude is
proportional to the energy
shift of the $\qqbar$ meson due to its interaction with nuclear
matter. The binding energy is given in terms of $\CM$ (neglecting
Fermi motion) by
\eqn\bindingdef{
 U_{{\rm binding}} = {\CM n\over 2 M},
}
where $n$ is the number of nucleons per unit volume in nuclear matter.
The factor of $2M$ divides out the standard relativistic normalisation
of nucleon states used in computing $\CM$ in Eq.~\fscat.
The binding energy of $\qqbar$ states in nuclear matter is
\eqn\binding{
U_{{\rm binding}} = \rho
 V_2(\rq)\left({\ce+\cb\over\rq^3}\right)\left( \gamma^2 -
\frac 14  \right) +
\rho \left({\ce-\cb\over\rq^3}\right){2\pi \over b_Q \alpha_s(\rq)},
}
where $\rho=Mn$ is the mass density of nuclear matter.

To determine the value of the binding energy of a $\qqbar$ state, we need
expressions for $\ce$ and $\cb$. The
coefficient $\cb$ is suppressed relative to $\ce$ by a factor of
$\alpha_s(\rq)^2$, and can be neglected.
The computation of $\ce$ is subtle, because gluon interactions
cause mixing between the singlet and octet $\qqbar$
channels.  In the heavy
quark limit, $m_Q\rightarrow\infty$, in which the $\qqbar$ system is
a Coulomb bound state,
a careful derivation
of $\ce$ has been given by Peskin \ref\peskin{M.E.~Peskin,
\np{156}{1979}
{365}.}
and we will quote his result.  For the $1s$ state, he finds
\eqn\aepeskin{
{\ce\over\rq^3}
%{1\over 2} \left({32\over N_c}\right)^2 \sqrt{\pi}
%\ {\Gamma(9/2)\over \Gamma(7)}\rqb^3
= {14\pi\over 27}\rqb^3}
%= {896\pi\over 729 \alpha^3(\rq) m_Q^3},
%}
where $N_c$ is the number of colours, and $\rqb$ is the Bohr radius
\eqn\bohrrad{\rqbinv={2\over 3}\alpha_s\left(\rqbinv\right) m_Q.}
Peskin used the $1/N_c$ expansion in obtaining the expression
for $\ce$ in Eq.~\aepeskin. This
approximation was used merely to give an expression which can be
written in a simple form. The error in not keeping the $1/N_c$ corrections
was estimated to be \peskin\  approximately 10\%, which is smaller
than the $\alpha_s(\rq)$ and $1/m_Q$ corrections that we have neglected.
Combining Eqs.~\aepeskin\ and \binding\ gives
\eqn\bindexp{
U_{{\rm binding}} = {14\over 27}\pi\rqb^3\rho
\left[ V_2(\rq)\left( \gamma^2 -
\frac 14  \right) +
{2\pi \over b_Q \alpha_s(\rq)}\right].
}

Treating the $\Upsilon$ as a Coulomb bound state, we find
from \bohrrad\ $\rqbinv\simeq 1.1\ \gev$.  Using a more realistic
potential, Quigg and Rosner \ref\qr{C.~Quigg and J. L.~Rosner,
\physrev{D23}{1981}{2625}.} calculated $\langle r \rangle_\Upsilon\simeq
1.0\ \gev^{-1}$, corresponding to a Bohr radius $\rqbinv\simeq{\frac 32}
\langle r \rangle^{-1}\simeq 1.5\ \gev$.  To the degree to which our
results are sensitive to the nonperturbative region of the potential,
they are not to be trusted, and since $U_{\rm binding}$ goes like $\rqb^3$,
this introduces a large uncertainty into our results.  We give an idea
of this uncertainty by giving $U_{\rm binding}$ for both values of
$\rqbinv$.
For an $\Upsilon$ or $\eta_b$, we find the
binding energies for the state at rest in nuclear matter of
\eqn\bbind{\eqalign{
U_{{\rm binding}} &\simeq 4\ \mev\ (\rqbinv=1.1\ \gev,\ \alpha_s=0.36),\cr
&\simeq 2\ \mev\ (\rqbinv=
 1.5\ \gev,\ \alpha_s=0.30).
}}
Here we have taken $\rho=1.2 \times 10^{-3}\ {\rm GeV}^4$, and
$b_Q=27/3$.

For the $J/\psi$, the validity of our expansion becomes markedly
more suspect:  from \bohrrad\ we find $\rqbinv\simeq 640\ \mev$, while
from \qr\ we find $\rqbinv\simeq 750\ \mev$, where we
really do not trust our perturbative results.
However, proceeding as before we find
\eqn\cbind{\eqalign{
U_{{\rm binding}} &\simeq 11\ \mev\ (\rqbinv=640\ \mev,\ \alpha_s=0.6),\cr
&\simeq 8\ \mev\ (\rqbinv=750\ \mev,\ \alpha_s=0.5).}}
Of course we do not believe this result is
anything more than a rough estimate for the $J/\psi$ binding energy,
as the
corrections to \cbind\ will be of order unity.

Note that our computation unambiguously
determines the sign of the energy shift of $\qqbar$ mesons at rest
in nuclear matter, at least in the $m_Q\rightarrow\infty$ limit.
The spin dependent corrections are suppressed relative to Eqs.~\bbind\
and \cbind\ by an additional factor of $\alpha_s$, so that the binding
energy of the $\eta_b$ and $\Upsilon$ are approximately equal. The same
arguments will also apply to the $J/\psi$ and $\eta_c$, but the spin
symmetry breaking effects will be larger.

There are two trivial corrections to the above expression
Eq.~\bindexp, which we have not included because they are
small compared to the $\alpha_s(\rq)$ and $1/m_Q$ terms that
we have neglected. The first is that there is a binding
energy correction to the relation between the nucleon mass
density, and the mass density of nuclear matter. The second
is that the structure functions of bulk nuclear matter
differ from those of nucleons, so that $\lambda$ in
Eq.~\bindexp\ should properly refer to the gluon
distribution measured using the $F_2$ structure function for
nuclear matter.

The results obtained in this paper show that the scattering
amplitude of a $\qqbar$ meson with nuclear matter scales as
$r_Q^3$, and the cross-section as $r_Q^6$ (up to the logarithmic
dependence in $\alpha_s$).\foot{The $r^3$ dependence of the two gauge
boson--meson scattering amplitude also shows that the Rayleigh
scattering cross-section is proportional to $r^6 \omega^4$. The nuclear
binding of $\qqbar$ mesons arises from the same operator that is
responsible for Rayleigh Scattering, with the electromagnetic field
strength tensor replaced by the gluon field strength tensor.}
This does not agree with the
assumption in the literature that the cross-section scales like the area
of the meson. We have also shown that the sign of the interaction energy
is negative, so that the meson binds to nuclear matter. There are no
pion corrections to our result, Eq.~\bindexp. All pion contributions
are included in the matrix element of the gluon operators in nuclear
matter. The only role of the pions in this calculation
is to modify the density and pressure of nuclear matter.
We emphasise that our results are rigorous
predictions of QCD and are completely model independent in the limit
$m_Q\rightarrow\infty$.

The largest source of error in our computation is the use
of Coulomb wavefunctions to compute $\ce$. Non-perturbative
confinement effects can be included by using a more realistic
potential model to compute the bound state wavefunction.
Unfortunately, the computation of $\ce$ involves intermediate
colour octet states, and thus the $\qqbar$  Hamiltonian in the colour
octet channel. In perturbation theory (the computation of Peskin \peskin),
one replaces the attractive $-4\alpha_s/3r$ attractive potential
in the singlet channel by a repulsive $\alpha_s/6r$ repulsive
potential in the octet channel. The corresponding modification
when non-perturbative effects are included in the singlet channel
is more complicated, and is being investigated further.

The binding energy was computed in Eq.~\bindexp\ for the $1s$ ground
state. One can compute the binding energies of the excited states
using the values of $\ce$ given in \peskin\ for the excited
states.   Since the binding energy goes like $r_Q^3$, the
excited states of quarkonium will be more tightly bound in the
$m_Q\rightarrow\infty$ limit.  Na\"\i vely applying $\bindexp$ to the
$\psi^\prime(2s)$ gives
\eqn\waybound{U_{\rm binding}^{(2s)} = \left({502\over 7}\right)
U_{\rm binding}^{(1s)}\sim 700\ \mev,}
suggesting that the $\psi^\prime(2s)$ has a lower
energy in nuclear matter than the ground state. The large radius of the
$2s$ is clearly an indication that the multipole
expansion is breaking down, so this result
cannot be trusted (but it might still be true).
\bigskip

We would like to thank S.~Brodsky, M.~Butler, D.~Wasson, R.~Milner,
M.~Wise
and D.~Kaplan for discussions.  M.~S. thanks the Institute for Nuclear
Theory at the University of Washington for their hospitality while some
of this work was done.  As this work was being completed, we
became aware of similar work in progress by J. Pasupathy\ref\pasu
{J.~Pasupathy, private communication.}.
This work was supported in part by DOE
grant \#DE-FG03-90ER40546, and by NSF grants PHY-8958081 and PHY-9057135.

\bigskip\bigskip
\listrefs
\end